# The Impact of a Raw Material Import Ban on Vertical Outward FDI: Theoretical Insights and Quasi-Experimental Evidence

Sajid Anwar
School of Business and Creative Industries
University of Sunshine Coast, Sippy Downs, QLD 4556, Australia
Email: SAnwar@usc.edu.au

Sizhong Sun
College of Business Law and Governance
James Cook University, Douglas, QLD 4814, Australia
Email: Sizhong.Sun@jcu.edu.au

**Abstract**

This paper examines how adverse supply-side shocks in domestic input markets influence firms' vertical outward foreign direct investment (OFDI) decisions. While the theoretical basis for cost-driven OFDI is well established, empirical evidence on the causal mechanisms remains limited. We develop a framework in which input cost shocks raise unit production costs, but firms undertake vertical OFDI only when shocks are sufficiently severe or when baseline costs are already high. Firm heterogeneity leads to a sorting pattern, whereby more productive firms are more likely to invest abroad. To test this mechanism, we exploit China's 2017 waste paper import ban as an exogenous shock and leverage a distinctive feature of the paper product industry's supply chain. Using a difference-in-differences strategy and firm-level data from 2000 to 2023, we find that the policy shock increased the probability of vertical OFDI by approximately 16% in the post-policy period relative to a control group. These results provide robust evidence that firms respond to domestic input shocks by reallocating production across borders, highlighting vertical OFDI as a strategic response to supply-side disruptions. The findings contribute to understanding the micro-foundations of global production decisions in the face of input market volatility.

* The authors thank Brian Copeland, Maria Ptashkina, Pascalis Raimondos, Joel Rodrigue, Yong Tan, and participants at the 34th Annual Conference of the Chinese Economics Society Australia and the 18th Australasian Trade Workshop for their valuable comments and suggestions. All remaining errors are the authors' sole responsibility.

# 1. Introduction

Firms invest overseas for various reasons, including the desire to serve foreign consumers and to seek for lower production costs (for a discussion, see Davies & Markusen, 2021). The motivation to serve foreign consumers typically leads to horizontal foreign direct investment (FDI).[1] In contrast, the pursuit of lower production costs drives vertical FDI, which is associated with vertical multinational enterprises (MNEs) that focus on optimizing efficiency through the geographical fragmentation of production processes. These distinct types of FDI have been the subject of extensive theoretical and empirical research, with considerable focus on understanding the firm-level determinants of FDI choices.

In the case of horizontal FDI, theoretical models initially assumed that firms are homogeneous, meaning all firms would respond similarly to incentives for investing abroad. Foundational works such as Markusen (1984), Horstmann and Markusen (1987, 1992), Brainard (1993) largely followed this framework, investigating FDI in the context of international trade. These studies were extended by Markusen and Venables (1997, 1998) and others, providing a stronger theoretical foundation for understanding the market-seeking behavior of MNEs. However, these representative/homogeneous firm models were unable to explain real-world observations that only a subset of firms, typically the most productive ones, engage in horizontal FDI. This gap was addressed by incorporating ex ante firm heterogeneity into trade and FDI models (Helpman, Melitz, & Yeaple, 2004; Luckstead, Devadoss, & Zhao, 2024), revealing that more productive firms engage in horizontal FDI, while less productive firms do not. This theoretical development has been crucial in reconciling empirical evidence with models of firm behavior in international markets (see for example Behrens & Picard, 2007; Egger & Pfaffermayr, 2005; Wang & Anwar, 2022).

Parallel to these developments, vertical FDI has also garnered significant scholarly attention. Early models, such as those by Helpman (1984), Helpman and Krugman (1985), Helpman (1985), Gao (1999), Markusen (2002), and Hanson, Mataloni and Slaughter (2003), examined the efficiency-seeking motivations of firms looking to reduce costs by offshoring stages of production. A key advancement in this field came with Antràs and Yeaple (2014), who introduced firm heterogeneity into a one-factor model of vertical FDI, demonstrating that more productive firms are more likely to engage in vertical FDI, much like the pattern observed for horizontal FDI. This sorting pattern underscores the central role of firm productivity in determining whether firms can benefit from fragmenting their production processes across countries.

While horizontal and vertical FDI have traditionally been studied separately, a third strand of research considers both horizontal and vertical FDI, often focusing on assessing which type is more compatible with empirical data (see for example Aizenman & Marion, 2004; Carr, Markusen, & Maskus, 2001; Davies, 2008; Markusen & Maskus, 2002; Neary, 2009; Ramondo, Rappoport, & Ruhl, 2012; Waldkirch, 2010; Zhao, 2001).[2] Existing studies frequently demonstrate stronger empirical support for horizontal FDI at the aggregate level, despite concerns about the difficulty of distinguishing between horizontal and vertical motivations for FDI (Lankhuizen, 2014). Nevertheless, this support for horizontal FDI does not diminish the role of vertical FDI in practice. Braconier, Norbäck and Urban (2005) emphasise that vertical FDI is crucial for understanding firm behavior, even if its empirical manifestations are more complex and context-dependent than those of horizontal FDI.

---

[1] It is synonymous with horizontal multinational enterprises (MNEs) that engage in market-seeking behavior.

[2] Comprehensive surveys of the relevant literature are provided by Markusen and Maskus (2003), Yeaple (2013), and Antràs and Yeaple (2014).

In this paper, we investigate vertical FDI from both theoretical and empirical perspectives, focusing on the effects of an exogenous shock induced by a raw material import ban. Our theoretical framework is grounded in a simple partial equilibrium model that incorporates firm heterogeneity within a monopolistically competitive product market. We demonstrate that the relative unit cost of intermediate inputs—comparing domestic costs to those of foreign inputs—and firm heterogeneity are crucial determinants influencing a firm's optimal decision to pursue vertical FDI.

Specifically, we posit that if the domestic unit cost of intermediate inputs is lower than the corresponding foreign cost, then vertical FDI will not be undertaken, regardless of firm productivity. Conversely, when domestic costs are comparatively higher, a sorting pattern emerges akin to that identified by Antràs and Yeaple (2014). To refine our model further, we endogenize the domestic unit cost by analyzing the upstream home intermediate input market, where supply is negatively impacted by the imposition of the raw material import ban. The equilibrium in this market links the import ban to the likelihood of engaging in vertical outward FDI, thereby facilitating the alignment of our theoretical model with empirical data.

To empirically assess the impact of the raw material import ban on the probability of vertical FDI, we draw upon data from the paper product manufacturing sector in China. Our identification strategy capitalizes on two distinctive features of the dataset: the exogeneity of the policy shock and a specific characteristic of the supply chain in the paper product industry. In July 2017, the Chinese government implemented a waste import ban (inclusive of waste paper) as part of an environmental regulation, aimed at mitigating pollution associated with the processing of imported waste materials (e.g., plastics and paper). This policy shock was not explicitly designed to influence firms' investment decisions and thus acted as a natural experiment that significantly impacted input costs.

The ban on import of waste paper, primarily utilized for producing pulp—a key intermediate input in the paper industry—results in increased production costs for paper manufacturing. A notable aspect of the supply chain in this sector is that waste paper pulp is exclusively employed for producing corrugated paper and container board, whereas other paper products are manufactured primarily using wood pulp. Consequently, firms engaged in producing corrugated paper and container board are directly affected by this shock, while those focusing on other products are less impacted. This distinction allows us to categorize firms into treated and control groups, and, coupled with the quasi-experimental nature of the policy shock, enables us to effectively identify treatment effects through a difference-in-differences (DID) research design.

These unique characteristics of our dataset provide a rare opportunity to investigate firms' decisions regarding vertical investment abroad, distinguishing our research from existing studies on vertical FDI. Our empirical findings indicate that the policy shock increases the probability of vertical OFDI among firms by approximately 16%, which is both statistically and economically significant. Thus, we contribute to the literature on vertical FDI by confirming a mechanism that drives such investment—namely, an increase in domestic input costs—which, while intuitive and conceptually recognized, has historically lacked robust empirical validation.

The remainder of the paper is structured as follows. Section 2 develops a theoretical model to examine how a supply-side shock in the upstream market—specifically, the raw material import ban—affects downstream firms' likelihood of undertaking vertical OFDI. Section 3 provides background on the 2017 waste paper import ban implemented by the Chinese government, which serves as the context for the policy shock. Section 4 outlines the DID research design and discusses the identification strategy. Section 5 describes the data and the construction of key variables, while Section 6 presents and interprets the DID estimation results. Section 7 concludes the paper.

## 2. Theoretical Model

Consider a supply chain that consists of three stages: raw materials, intermediate inputs and final goods, as illustrated in Figure 1. A shock[3] occurs in the upstream raw material market, reducing the supply of raw materials used to produce intermediate inputs. How does this supply-side shock affect firms in the downstream final goods market with respect to overseas vertical investment? We model this situation, and then empirically test the prediction within the paper manufacturing supply chain.

Figure 1. The Supply Chain

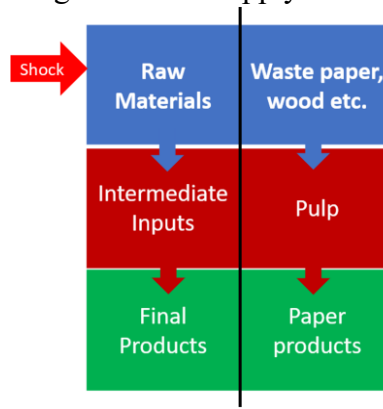

The final goods market is monopolistically competitive, with consumers exhibiting preferences represented by a constant elasticity of substitution (CES) utility function. Maximizing this utility function subject to a budget constraint yields the following demand function:

$$q = A p^{\frac{1}{\rho-1}}$$

where $q$ and $p$ represent the quantity demanded and price of a variety, respectively; $\rho$ is the CES preference parameter with $\rho \in (0,1)$; and $A$ is the demand shifter (income divided by a price index), which firms take as given since they are small relative to the market.

Given this demand function, firms decide on pricing and whether to invest in a foreign market by establishing a subsidiary for producing intermediate inputs (i.e., vertical outward

---

[3] We consider long-standing/permanent, rather than transitory, shocks.

FDI).[4] Firms are heterogeneous in the Melitz (2003) sense, with productivity endowment $\lambda$ following a Pareto distribution, characterized by a scale parameter $\lambda_m > 0$ and shape parameter $\alpha$ ($\alpha > \frac{1-\rho}{\rho}$). Production incurs a fixed cost $f$, reflecting investments in facilities such as assembly lines. With these facilities, one worker utilizes $\eta$ units of intermediate inputs to produce $\lambda$ units of output.[5]

The wage rate is normalized to 1 by choosing an appropriate numeraire. The unit cost of intermediate inputs is $\delta$ when sourced domestically, and $\tilde{\delta}$ if produced and imported from the firm's foreign subsidiary.[6] A firm that undertakes vertical OFDI incurs a fixed investment cost, $f_I$.[7] Let $\chi_I = 1$, if the firm engages in vertical OFDI, and $\chi_I = 0$ otherwise. The firm's marginal cost of production, $c$, is given by:

$$c = \frac{1 + \eta[(1-\chi_I)\delta + \chi_I \tilde{\delta}]}{\lambda}$$

Accordingly, the firm's profit is:

$$\pi = (p - c)Ap^{\frac{1}{\rho-1}} - f - f_I \chi_I$$

Firms set prices by charging a markup over marginal cost, such that $p = \frac{c}{\rho}$, yielding the following optimal profit:

$$\pi = (1-\rho)\rho^{\frac{\rho}{\rho-1}}A\{1 + \eta[(1-\chi_I)\delta + \chi_I \tilde{\delta}]\}^{\frac{\rho}{\rho-1}}\lambda^{\frac{\rho}{1-\rho}} - f - f_I \chi_I \tag{1}$$

*2.1 Optimal Decision on Vertical OFDI*

Let $\pi(\chi_I = 1)$ and $\pi(\chi_I = 0)$ denote the firm's profit with and without vertical OFDI, respectively:

$$\pi(\chi_I = 1) = (1-\rho)\rho^{\frac{\rho}{\rho-1}}A(1 + \eta\tilde{\delta})^{\frac{\rho}{\rho-1}}\lambda^{\frac{\rho}{1-\rho}} - (f + f_I)$$

$$\pi(\chi_I = 0) = (1-\rho)\rho^{\frac{\rho}{\rho-1}}A(1 + \eta\delta)^{\frac{\rho}{\rho-1}}\lambda^{\frac{\rho}{1-\rho}} - f$$

The firm's decision to undertake vertical OFDI depends on the relative magnitude of $\pi(\chi_I = 1)$ and $\pi(\chi_I = 0)$. If both profits are negative, the firm will not enter the market, resulting in an endogenous self-selection. Otherwise, the firm will invest abroad if $\pi(\chi_I = 1) > \pi(\chi_I = 0)$.

A closer examination of the profit functions reveals that the relative magnitude depends on two key factors: $\delta/\tilde{\delta}$ and $\lambda$. Both profit functions are increasing in the firm's productivity endowment ($\lambda$), with $\lim_{\lambda \to 0} \pi(\chi_I = 0) = -f$ and $\lim_{\lambda \to 0} \pi(\chi_I = 1) = -(f + f_I)$. If $\delta/\tilde{\delta} \leq 1$, namely intermediate inputs are cheaper domestically, then the slope of $\pi(\chi_I = 0)$ with respect to $\lambda^{\frac{\rho}{1-\rho}}$ exceeds that of $\pi(\chi_I = 1)$. Since $\lim_{\lambda \to 0} \pi(\chi_I = 0) > \lim_{\lambda \to 0} \pi(\chi_I = 1)$, firms will not engage in vertical outward FDI upon market entry, as illustrated in Figure 2.

---

[4] Alternatively, firms may consider undertaking mergers and acquisitions (M&A) in the foreign market.

[5] We implicitly assume domestic and foreign intermediate inputs are perfect substitute for each other. Alternatively, one could model intermediate inputs as a CES aggregation of domestic and foreign varieties.

[6] The parameter $\tilde{\delta}$ encompasses all additional import-related costs. Note that we continue to assume domestic and foreign intermediate inputs are perfect substitutes.

[7] For M&A, $f_I$ represents the fixed cost associated with the M&A process.

Figure 2. No Vertical Outward FDI ($\delta \leq \tilde{\delta}$)

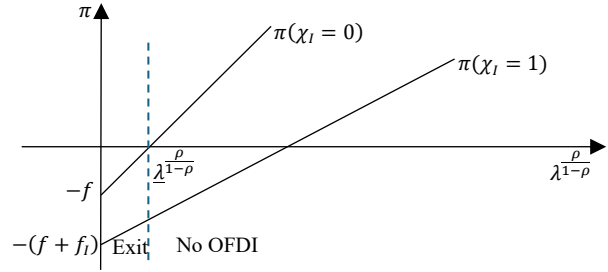

Conversely, if $\delta/\tilde{\delta} > 1$, intermediate inputs are less expensive when sourced from a foreign subsidiary. More capable firms (those with higher $\lambda$) derive greater benefit from lower input costs abroad and, as a result, are more likely to invest overseas. The fact that $\lim_{\lambda \to 0} \pi(\chi_I = 0) > \lim_{\lambda \to 0} \pi(\chi_I = 1)$, combined with both profit functions being upward sloping, with $\pi(\chi_I = 1)$ exhibiting a steeper slope, produces a sorting pattern. Specifically, less productive firms do not enter the market, moderately productive firms serve the market without vertical OFDI, and the most productive firms engage in vertical OFDI. This sorting pattern is illustrated in Figure 3 and is consistent with the findings of Antràs and Yeaple (2014).[8]

Figure 3. Vertical Outward FDI ($\delta > \tilde{\delta}$)

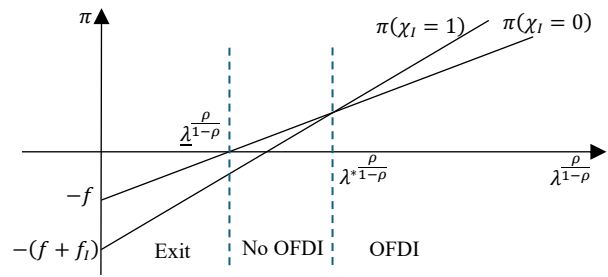

In Figures 2 and 3, $\underline{\lambda}$ represents the cut-off productivity endowment, below which firms earn negative profit (and thus do not enter the market). Using the profit function $\pi(\chi_I = 0)$, we can solve for $\underline{\lambda}$ as follows:

$$\underline{\lambda} = (1-\rho)^{-\frac{1-\rho}{\rho}} \rho^{-1} A^{-\frac{1-\rho}{\rho}} (1+\eta\delta) f^{\frac{1-\rho}{\rho}} \qquad (2)$$

For vertical OFDI, the condition that $\pi(\chi_I = 1) = \pi(\chi_I = 0)$ defines a cut-off productivity endowment ($\lambda^*$), as follows:

---

[8] While Antràs and Yeaple's (2014) focus primarily on the role labor, our analysis explicitly incorporates intermediate inputs, allowing us to examine the equilibrium outcomes within the intermediate input market—a dimension that has not been previously unexplored.

$$\lambda^* = (1-\rho)^{-\frac{1-\rho}{\rho}} \rho^{-1} A^{-\frac{1-\rho}{\rho}} \left[ (1+\eta\tilde{\delta})^{\frac{\rho}{\rho-1}} - (1+\eta\delta)^{\frac{\rho}{\rho-1}} \right]^{-\frac{1-\rho}{\rho}} f^{\frac{1-\rho}{\rho}} \quad (3)$$

We summarize these findings in the following proposition:

**Proposition 1**: Firms' optimal decision to undertake vertical outward FDI depends on two factors: the relative unit cost of intermediate inputs ($\delta/\tilde{\delta}$) and their productivity endowment ($\lambda$), as follows:

If $\delta/\tilde{\delta} \leq 1$: no firm will engage in vertical outward FDI, regardless of productivity.

If $\delta/\tilde{\delta} > 1$:

$\lambda \in (0, \underline{\lambda})$: firms do not enter the market.

$\lambda \in [\underline{\lambda}, \lambda^*]$: firms enter the market without vertical outward FDI.

$\lambda \in (\lambda^*, +\infty)$: firms enter the market and engage in vertical outward FDI.

*2.2 The Marginal Effect of Intermediate Input Cost on the Probability of OFDI*

Proposition 1 suggests that $\chi_I$ is a function of $\delta/\tilde{\delta}$ and $\lambda$, where $\lambda$ follows a Pareto distribution. By taking expectation with respect to $\lambda$, conditional on $\lambda \geq \underline{\lambda}$, we can derive the probability that a firm engages in vertical OFDI, as follows:

$$\mathbb{P}(\chi_I = 1) = \mathbb{1}(\delta/\tilde{\delta} > 1)\mathbb{P}(\lambda > \lambda^* | \lambda \geq \underline{\lambda}) = \mathbb{1}(\delta/\tilde{\delta} > 1) \frac{\left(\frac{\lambda_m}{\lambda^*}\right)^\alpha}{\left(\frac{\lambda_m}{\underline{\lambda}}\right)^\alpha}$$

$$= \mathbb{1}(\delta/\tilde{\delta} > 1) \left[ \left(\frac{1+\eta\tilde{\delta}}{1+\eta\delta}\right)^{\frac{\rho}{\rho-1}} - 1 \right]^{\frac{\alpha(1-\rho)}{\rho}} \left(\frac{f}{f_I}\right)^{\frac{\alpha(1-\rho)}{\rho}} \quad (4)$$

where $\mathbb{P}$ denotes probability, $\mathbb{1}(\cdot)$ is an indicator function that takes the value of 1 if the argument is true, and the second equality follows from the Pareto distribution of $\lambda$. The third equality is obtained by plugging in equations (2) and (3). Additionally, we have $\delta \leq \left(1+\frac{f_I}{f}\right)^{\frac{1-\rho}{\rho}} \frac{1+\eta\tilde{\delta}}{\eta} - \frac{1}{\eta}$. For $\delta > \left(1+\frac{f_I}{f}\right)^{\frac{1-\rho}{\rho}} \frac{1+\eta\tilde{\delta}}{\eta} - \frac{1}{\eta}$, $\mathbb{P}(\chi_I = 1) = 1$. It is important to condition on firms entering the market (i.e., $\lambda \geq \underline{\lambda}$), which captures firms' endogenous self-selection into the market (and the data).

Figure 4 presents the graph of $\mathbb{P}(\chi_I = 1)$, where $(\alpha, \rho, \tilde{\delta}, f, f_I) = (2, 0.5, 2, 1, 1)$ and $\eta \in \{1, 2, 10\}$ (representing low, medium, and high intensity of intermediate input usage, respectively). Three patterns emerge from this graph.

First, when the unit cost of intermediate inputs is low, firms do not engage in vertical OFDI, as outlined in Proposition 1. Second, as the domestic cost of intermediate inputs ($\delta$) rises, the probability that firms invest overseas increases, eventually reaching 1. Third, firms that utilize intermediate inputs more intensively are more likely to invest overseas. By partially differentiating equation (4) with respect to $\eta$, we find that the marginal effect is positive. The second and third patterns underscore the crucial role that input cost considerations play in firms' decisions to invest abroad.

Figure 4. Probability of Vertical Outward FDI

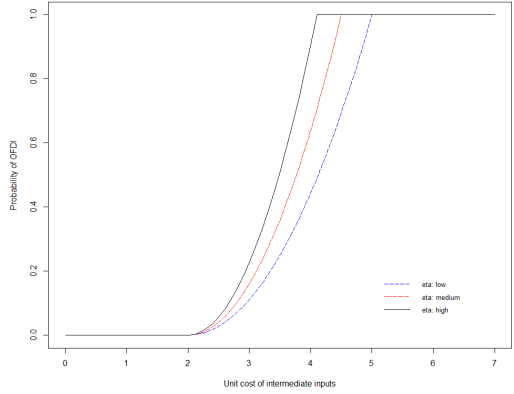

Later, we will empirically examine how an exogenous policy shock (e.g., China's waste paper import ban) that raises the unit cost of intermediate inputs influences firms' investment decisions. A theoretical prediction of this marginal effect provides a foundation for bringing the model to the data.

When the unit cost in the domestic market is relatively low ($\delta/\tilde{\delta} \leq 1$), the probability of investment, $\mathbb{P}(\chi_I = 1)$, is zero, and thus $\frac{\partial \mathbb{P}(\chi_I=1)}{\partial \delta} = 0$. In contrast, when $\delta/\tilde{\delta} > 1$, partially differentiating equation (4) with respect to $\delta$ yields the following marginal effect:

$$\frac{\partial \mathbb{P}(\chi_I = 1)}{\partial \delta} =$$

$$\mathbb{1}(\delta/\tilde{\delta} > 1)\alpha \left(\frac{f}{f_I}\right)^{\frac{\alpha(1-\rho)}{\rho}} \left[\left(\frac{1+\eta\tilde{\delta}}{1+\eta\delta}\right)^{\frac{\rho}{\rho-1}} - 1\right]^{\frac{\alpha(1-\rho)}{\rho}-1} \left(\frac{1+\eta\tilde{\delta}}{1+\eta\delta}\right)^{\frac{\rho}{\rho-1}-1} \frac{1+\eta\tilde{\delta}}{(1+\eta\delta)^2}\eta > 0$$

This indicates that an increase in the unit cost of intermediate inputs in the domestic market induces firms to invest overseas, all else being equal.

However, when $\delta$ crosses the threshold $\tilde{\delta}$, standard differentiation cannot be applied due to the kink introduced by $\mathbb{1}(\delta/\tilde{\delta} > 1)$. In this case, if we let $\delta$ increase from $\delta_L \leq \tilde{\delta}$ to $\delta_H > \tilde{\delta}$, then the discrete change (i.e., $\Delta$) in the probability of vertical OFDI is: $\Delta\mathbb{P}(\chi_I = 1) \equiv$

$\mathbb{P}(\chi_I = 1|\delta_H) - \mathbb{P}(\chi_I = 1|\delta_L) = \left[\left(\frac{1+\eta\tilde{\delta}}{1+\eta\delta_H}\right)^{\frac{\rho}{\rho-1}} - 1\right]^{\frac{\alpha(1-\rho)}{\rho}} \left(\frac{f}{f_I}\right)^{\frac{\alpha(1-\rho)}{\rho}} > 0$. This implies that a

discrete increase in $\delta$ from below $\tilde{\delta}$ to above $\tilde{\delta}$ increases the probability of vertical OFDI. Intuitively, as domestic costs rise, the marginal benefit of investing overseas increases. Additionally, this benefit is larger for firms that use intermediate inputs more intensively (i.e., firms with a higher $\eta$). To confirm this, we can partially differentiate $\Delta\mathbb{P}(\chi_I = 1)$ with respect to $\eta$, which yields:

$$\frac{\partial \Delta \mathbb{P}(\chi_I = 1)}{\partial \eta} = \alpha \left(\frac{f}{f_I}\right)^{\frac{\alpha(1-\rho)}{\rho}} \left[\left(\frac{1+\eta\tilde{\delta}}{1+\eta\delta_H}\right)^{\frac{\rho}{\rho-1}} - 1\right]^{\frac{\alpha(1-\rho)}{\rho}-1} \left(\frac{1+\eta\tilde{\delta}}{1+\eta\delta_H}\right)^{\frac{\rho}{\rho-1}-1} \frac{\delta_H - \tilde{\delta}}{(1+\eta\delta_H)^2}$$
$$> 0$$

We summarize these results in Proposition 2 as follows:

**Proposition 2:** The marginal effect of intermediate input costs on the probability of vertical outward FDI is characterized as follows:

(i) If $\delta/\tilde{\delta} \leq 1$, the marginal effect is zero.

(ii) If $\delta/\tilde{\delta} > 1$, the marginal effect is positive.

(iii) A discrete increase in $\delta$ from below $\tilde{\delta}$ to above $\tilde{\delta}$ raises the probability of vertical outward FDI, particularly for firms with higher intensity of intermediate input use (i.e., larger $\eta$).

*2.3 Equilibrium in the Intermediate Input Market*

Thus far, $\delta$ has been treated as an exogenous parameter. We now endogenize $\delta$ by modelling the upstream domestic intermediate input market, assuming perfect competition. We examine the market equilibrium and analyze the comparative statics in response to an exogenous supply-side policy shock.

*2.3.1 Aggregate Supply*

The supply of intermediate inputs is represented by the following function:

$$M^S = M(\delta, \chi_m)$$

where $M^S$ denotes the aggregate supply of intermediate inputs, $\delta$ is the unit cost (or price), and $\chi_m$ is a supply-side policy variable. This variable takes the value 1 if imports of raw materials used to manufacture intermediate inputs are allowed, and 0 if not. The supply function $M^S$ is monotonically increasing in $\delta$ ($\frac{\partial M^S}{\partial \delta} > 0$). Moreover, $M(\delta, 1) \geq M(\delta, 0)$, namely a ban on raw material imports reduces the supply of intermediate inputs at any price level. Additionally, when the price of intermediate inputs drops to zero, the supply also falls to zero, regardless of the raw material import policy, i.e., $M(0,1) = M(0,0) = 0$. These assumptions imply that a raw material import ban causes the supply curve to rotate counterclockwise.

*2.3.2 Aggregate Demand*

The production technology determines an individual firm's optimal use of intermediate inputs as: $\rho^{\frac{1}{1-\rho}} \eta A \{1 + \eta[(1-\chi_I)\delta + \chi_I\tilde{\delta}]\}^{\frac{1}{\rho-1}} \lambda^{\frac{\rho}{1-\rho}}$, which depends on the intensity of intermediate input use ($\eta$) and its productivity endowment ($\lambda$), both of which vary across firms. Additionally, the fixed cost of production ($f$), which also varies across firms, affects the cut-off productivity (equations 2 and 3) and thus the aggregate demand for intermediate inputs.

To derive the aggregate demand, we proceed in two steps. First, for a given pair $(\eta, f)$, we integrate the intermediate input demand over all firms with productivity $\lambda$ above the cut-off level. Then, we average the resulting expression over the joint distribution of $(\eta, f)$ across the population of firms.

If $\delta \leq \tilde{\delta}$, given $(\eta, f)$, aggregating over $\lambda$, we obtain the demand for intermediate inputs as:
$M_1^D(\delta, \eta, f) = \frac{\alpha \rho^{1+\alpha} \lambda_m^\alpha}{\alpha - \rho/(1-\rho)} (1-\rho)^{\frac{\alpha(1-\rho)}{\rho} - 1} A^{\frac{\alpha(1-\rho)}{\rho}} (1+\eta\delta)^{-1-\alpha} \eta f^{1-\frac{\alpha(1-\rho)}{\rho}}$. Then averaging over $(\eta, f)$, we obtain: $M_1^D(\delta) = \frac{\alpha \rho^{1+\alpha} \lambda_m^\alpha}{\alpha - \rho/(1-\rho)} (1-\rho)^{\frac{\alpha(1-\rho)}{\rho} - 1} A^{\frac{\alpha(1-\rho)}{\rho}} \int (1+\eta\delta)^{-1-\alpha} \eta f^{1-\frac{\alpha(1-\rho)}{\rho}} dG(\eta, f)$, where $G(\cdot)$ is the joint cumulative distribution function of (CDF) of $(\eta, f)$. $M_1^D(\delta)$ is a monotonically decreasing function of $\delta$.

If $\delta > \tilde{\delta}$, only firms with $\lambda \in [\underline{\lambda}, \lambda^*]$ continue to source the intermediate inputs from the domestic market. Hence, aggregating over $\lambda$ given firm characteristics $(\eta, f, f_I)$, the aggregate demand for intermediate inputs is:

$$M_2^D(\delta, \eta, f, f_I) = \frac{\alpha \rho^{1+\alpha} \lambda_m^\alpha}{\alpha - \rho/(1-\rho)} (1-\rho)^{\frac{\alpha(1-\rho)}{\rho} - 1} A^{\frac{\alpha(1-\rho)}{\rho}} (1+\eta\delta)^{-1-\alpha} \eta f^{1-\frac{\alpha(1-\rho)}{\rho}} \left\{ 1 - \left[\left(\frac{1+\eta\tilde{\delta}}{1+\eta\delta}\right)^{\rho/(\rho-1)} - 1\right]^{\frac{\alpha(1-\rho)}{\rho} - 1} \left(\frac{f_I}{f}\right)^{1-\frac{\alpha(1-\rho)}{\rho}} \right\}$$

Averaging over $(\eta, f, f_I)$, we have:

$$M_2^D(\delta) = \frac{\alpha \rho^{1+\alpha} \lambda_m^\alpha}{\alpha - \rho/(1-\rho)} (1-\rho)^{\frac{\alpha(1-\rho)}{\rho} - 1} A^{\frac{\alpha(1-\rho)}{\rho}} \int (1+\eta\delta)^{-1-\alpha} \eta f^{1-\frac{\alpha(1-\rho)}{\rho}} \left\{ 1 - \left[\left(\frac{1+\eta\tilde{\delta}}{1+\eta\delta}\right)^{\rho/(\rho-1)} - 1\right]^{\frac{\alpha(1-\rho)}{\rho} - 1} \left(\frac{f_I}{f}\right)^{1-\frac{\alpha(1-\rho)}{\rho}} \right\} d\tilde{G}(\eta, f, f_I)$$

where $\tilde{G}(\cdot)$ is the joint CDF of $(\eta, f, f_I)$. $M_2^D(\delta)$ is also a monotonically decreasing function of $\delta$.

Consequently, the aggregate demand for intermediate inputs is:
$$M^D(\delta) = \mathbb{1}(\delta/\tilde{\delta} \leq 1) M_1^D(\delta) + \mathbb{1}(\delta/\tilde{\delta} > 1) M_2^D(\delta)$$

The properties of aggregate demand are as follows:

(i) $M^D(\delta)$ is a monotonically decreasing function of $\delta$ (Law of Demand).
(ii) $\lim_{\delta \to \tilde{\delta}^+} M_2^D(\delta) = M_1^D(\tilde{\delta})$, implying continuity of the demand curve with a kink at $\delta = \tilde{\delta}$.
(iii) $M^D(0) = M_1^D(0)$, and $\lim_{\delta \to \infty} M^D(\delta) = 0$.
(iv) When $\delta > \tilde{\delta}$, $\frac{\partial M_2^D}{\partial \delta} > \frac{\partial M_1^D}{\partial \delta}$, reflecting a steeper slope.

These properties allow us to draw the demand curve.

Figure 5 illustrates the supply-demand equilibrium in the domestic intermediate input market. It shows four aggregate supply curves corresponding to three distinct regimes under a supply-side shock, with each intersection representing a market clearing equilibrium. A supply shock—such as a raw material import ban—causes the supply curve to rotate counterclockwise, leading to an unambiguous increase in the equilibrium price (unit cost) of intermediate inputs. However, the implications of this cost increase for firms' decisions to engage in vertical outward FDI differ across the three regimes.

Figure 5. Equilibrium in the Intermediate Input Market

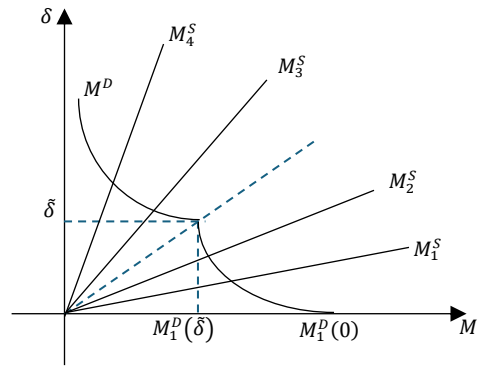

*Regime 1:* The supply curve rotates from $M_1^S$ to $M_2^S$, but the equilibrium unit cost remains below the threshold $\tilde{\delta}$. In this case, firms' probability of vertical OFDI remains zero. Consequently, the supply-side shock—such as a raw material import ban—has no impact on their decision to invest overseas.

*Regime 2:* The supply curve rotates from $M_2^S$ to $M_3^S$, causing the equilibrium unit cost to increase from below to above $\tilde{\delta}$. This rise in cost induces some firms—particularly those that are more productive and rely more heavily on intermediate inputs—to undertake vertical integration overseas.

*Regime 3:* The supply curve rotates from $M_3^S$ to $M_4^S$, with the equilibrium unit cost already above $\tilde{\delta}$ before the shock. In this case, since $\mathbb{P}(\chi_I = 1)$ is a monotonically non-decreasing function of $\delta$, any further increase in cost raises the likelihood of vertical outward FDI, unless the probability was already one before the shock.

The results concerning the impact of a raw material import ban on firms' probability of vertical OFDI can be summarized in a proposition as follows:

**Proposition 3:** Consider a raw material import ban that increases the domestic unit cost of intermediate inputs ($\delta$). The effect on firms' vertical OFDI decisions depends on the initial cost of intermediate inputs: (i) if the cost increase is modest such that $\delta$ remains below $\tilde{\delta}$, the ban has no effect on firms' decisions to invest overseas, (ii) if $\delta$ rises from below to above $\tilde{\delta}$, the ban induces some firms to invest overseas, and (iii) if $\delta$ is already above $\tilde{\delta}$, the ban further increases the probability of vertical OFDI, unless it is already 1.

While firms can invest overseas and import intermediate inputs from their foreign affiliates, another option is to directly import these inputs from external suppliers. In our theoretical model, we abstract away from this consideration since our focus is on vertical OFDI.

Introducing direct import as an option would not alter the main insights but would enrich the sorting structure of firms. Specifically, under the condition that $\delta/\tilde{\delta} > 1$ and assuming that the fixed cost of direct import is lower than that of vertical OFDI, the equilibrium sorting pattern would be as follows: the least productive firms do not enter the market; less productive firms serve the domestic market without importing or investing abroad; moderately productive firms import the intermediate inputs directly; and the most productive firms engage in vertical OFDI.

The next section empirically tests the model's predictions, using China's waste paper import ban as a quasi-natural experiment.

**3. Paper Manufacturing and Waste Paper Import Ban**

Pulp is a key intermediate input in the paper manufacturing industry, derived from either wood[9] or recycled waste paper. Wood pulp is primarily used to produce household paper, coated paper, double offset paper, and ivory board, which serve various purposes, including printing, packaging, and consumer use. In contrast, waste paper pulp is used to produce corrugated paper and container board, essential for carton manufacturing. This differentiation highlights the distinct supply chains within the paper industry.

Figure 6. The Supply Chain in the Paper Product Industry

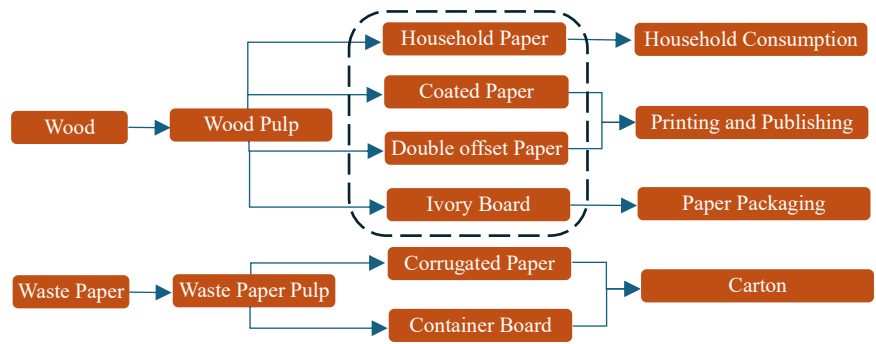

Note: The figure does not include a smaller proportion of non-wood pulp.
Source: Adapted from Figure 1 of
https://www.esmchina.com/marketnews/34401.html (in Chinese).

In the supply chain of the paper product industry (Figure 6), two distinct types of intermediate inputs—wood pulp and waste paper pulp—are produced from different raw materials and serve different purposes. Not all firms within the industry manufacture every type of paper products; some may specialize in products derived from wood pulp, while others focus on products made from waste paper pulp. Consequently, when the cost of one type of intermediate inputs rises, not all firms are equally affected. This variation provides a natural opportunity to identify the impact of an intermediate input cost shock.

One key source of raw material for producing waste paper pulp in China has been imports. Since the 1980s, China has imported waste materials, which were processed into industrial raw materials, including pulp for paper production. With its rapid economic growth, China's waste imports increased dramatically—from 0.45 million tons in 1995 to 4.85 million tons in 2016 (Shi & Zhang, 2023). By 2016, China had become the largest global importer of waste, accounting for 56% of the world's plastic waste exports (Brooks, Wang, & Jambeck, 2018). In 2017, China was the largest waste paper importer, accounting for 40% of the global waste paper trade (China Paper Association, 2019).

---
[9] Or other plants, such as the straw pulp and bagasse pulp.

Although waste imports contributed to China's economic growth by supplying raw materials for industrial production, they also imposed significant environmental costs. The processing of imported waste, including waste paper, often generates pollution if not properly managed. For instance, imported waste paper contains non-recyclable sundries (up to 1.5%, as per customs regulations), which are frequently incinerated without appropriate abatement measures (Nkwachukwu, Chima, Ikenna, & Albert, 2013). Furthermore, transforming waste paper into pulp emits pollutants, often untreated, which contain harmful substances (Huang, Zhou, Hong, Feng, & Tao, 2013).

The deteriorating environmental conditions—driven partly by waste imports—prompted the Chinese government to tighten its environmental regulations. In July 2017, the State Council introduced the *Implementation Plan for Banning the Entry of Foreign Waste and Promoting the Reform of the Management System of Solid Waste Import* (Decree No. 70, 2017).[10] This plan aimed to ban the import of certain pollution-intensive waste materials, including plastics, paper, and textiles, by the end of 2017. The ban was subsequently expanded to cover additional categories of waste in 2018.

As a result, China's waste imports declined sharply, falling from 42 million tons in 2017 to 22 million tons in 2018 and 13 million tons in 2019 (Shi & Zhang, 2023). Existing studies suggest that this policy has significantly reduced pollution. For instance, Shi and Zhang found that prefectures with higher levels of waste imports before the policy saw greater improvements in air quality post-policy. Similarly, Unfried and Wang (2024) reported improvements in air quality following reductions in plastic waste imports.

While the waste import ban was primarily aimed at environmental protection, it also had unintended economic consequences, particularly for industries reliant on waste imports. In the paper product industry, the loss of imported waste paper as a source of raw material led to an increase of the price of domestic waste paper[11] and a reduction of domestic waste paper pulp supply. The annual domestic production of waste paper pulp decreased from almost 64 million tons in 2015-2017 to less than 54 million tons in 2018-2020.[12] Consequently, the cost of waste paper pulp, a crucial intermediate input for paper manufacturing, increased, which pressured the downstream paper manufacturing firms (United Credit Ratings, 2018).[13]

According to our theoretical model, firms may respond to this increase in input costs by investing overseas in the upstream production process through vertical OFDI, should the cost rise be substantial. Indeed, a number of news reports/analysis have found that paper manufacturing firms have sought to invest overseas in the upstream sub-industry, in order to mitigate negative effect of the policy shock.[14]

---

[10] Details regarding this policy are available at https://www.gov.cn/zhengce/content/2017-07/27/content_5213738.htm (in Chinese).

[11] See https://qiye.chinadaily.com.cn/a/202012/29/WS5fea8df3a3101e7ce9737f81.html.

[12] See http://www.21jingji.com/article/20210601/herald/034a37498dd780bc16eee82fb8a04d5c.html (in Chinese).

[13] At the very least, it is perceived to contribute to higher costs of waste paper pulp.

The waste import ban represents an exogenous shock to firms' overseas investment decisions. The Chinese government did not introduce the policy to either encourage or discourage FDI, making the policy shock quasi-experimental in nature. Moreover, as illustrated in Figure 6, the ban affected firms within the same industry to varying degrees. Firms that rely on waste paper pulp were more significantly impacted, while those that used wood pulp were less affected. This distinction enables the identification of treated and control firms, allowing us to estimate the policy's impact on firms' vertical OFDI decisions using a DID research design.

## 4. Estimation Strategy

To estimate the treatment effect of China's waste paper import ban on firms' likelihood of engaging in vertical OFDI, we recognize the quasi-experimental nature of the policy, as outlined in Section 3. Furthermore, the distinct characteristics of the paper product manufacturing supply chain enable us to distinguish between treated and control firms. This allows for the implementation of a DID research design to assess the policy's impact. Below, we detail our empirical strategy at both the disaggregated firm level and the aggregate level.

*4.1 Firm level analysis*

*4.1.1 Research Design*

We employ the following regression model to estimate the treatment effect of waste paper import ban on firms' probability of conducting vertical outward FDI:[15]

$$OFDI_{ijt} = \beta_0 + \beta_1 dP_t + \beta_2 dP_t \times dT_j + x'_{ijt}\beta_3 + v_{ij} + \varepsilon_{ijt} \qquad (5)$$

where the outcome variable $OFDI$ is a cumulative measure, which takes a value of 1 if firm $i$ in group $j \in \{0,1\}$ has ever engaged in vertical OFDI up to and including year $t$, and 0 otherwise; $dP$ is a dummy variable, which takes a value of 1 in the post-policy period (starting in 2017);[16] $dT$ denotes the treatment status, taking a value of 1 for treated firms ( $j = 1$, those producing corrugated paper and container board) and 0 otherwise ($j = 0$); $x_{ijt}$ is a vector of control variables that conceptually affect firms' overseas investment decisions; $v_{ij}$ is the firm ($v_i$) and group ($v_j$) fixed effect, namely $v_{ij} = v_i + v_j$ with $E[v_i] = E[v_1]$;[17] and $\varepsilon_{ijt}$ is the error term. The parameter $\beta_2$ captures the treatment effect. If the waste paper import ban significantly raises the price of waste paper pulp, we expect an increase in the probability that firms will engage in vertical outward FDI, as predicted by the theoretical model. Consequently, we expect the estimate of $\beta_2$ to be significantly positive.

---

[14] For example, see https://m.21jingji.com/article/20210524/herald/b3c81f99e00e1e2df31698a4e5f12c8f.html; https://m.huxiu.com/article/442175.html; https://finance.sina.cn/hkstock/ggyw/2020-10-31/detail-iiznctkc8667787.d.html?from=wap; https://www.163.com/dy/article/HEJ60E26055360RU.html (in Chinese).

[15] Equation (5) is a linear probability model. In addition, one may wish to estimate corresponding probit and logit models. Nevertheless, as exhibited in Figure 7, firms in the control group do not conduct vertical OFDI in the sample period ($OFDI = 0$). This lack of variation renders estimating probit and logit models infeasible.

[16] Note that the definition includes the year 2017. We include 2017 because firms' decisions regarding vertical outward FDI are likely driven by their perception, rather than the actual increase, in the cost of intermediate inputs. Our results remain consistent even when 2017 is excluded from the analysis.

[17] Note $E[v_i|x_{ijt}, dP_t, dT_j] = E[v_i|x_{ijt}, dT_j]$, where the equality is due to exogeneity of the policy shock, can depend on $x_{ijt}$ and $dT_j$. Then, one can apply the law of iterated expectation to obtain $E[v_i] = E\left[E[v_i|x_{ijt}, dP_t, dT_j]\right]$, which we assume is i.i.d. over firms and does not depend on $j$.

A note on the definition of the outcome variable $OFDI_{ijt}$ is warranted here. Overseas investments aimed at securing cheaper intermediate inputs are often one-off. That is, once a firm has made such a vertical OFDI, it may not need to make additional investments unless market conditions change. This distinguishes vertical OFDI from horizontal OFDI, which seeks to serve consumers in different geographic regions. The definition of $OFDI_{it}$ aligns with this feature of vertical OFDI.

*4.1.2 Identification*

Identification of $\beta_2$ critically depends on the assumption that $E[\varepsilon_{ijt}|dP_t, dT_j, x_{ijt}] = 0$. While the policy shock $dP_t$ is exogenous, one might argue that more (or less) productive firms choose to produce corrugated paper and container board.[18] Since more productive firms are more likely to invest overseas, $dT_j$ is then correlated with the unobserved firm productivity. In the DID setup, we use the firm fixed effect $\nu_{ij}$ to control such a possibility. The within-firm demeaning in the Fixed Effect (FE) estimation will eliminate $\nu_{ij}$, hence addressing the possible endogeneity of $dT_j$. The control variables may also be endogenous. If this is the case, $E[\varepsilon_{ijt}|dP_t, dT_j, x_{ijt}] = g(x_{ijt})$. If $g(x_{ijt})$ is linear, it will also be absorbed into $x'_{ijt}\beta_3$. If $g(x_{ijt})$ is non-linear, we can include polynomial functions of $x_{ijt}$ to approximate it, assuming it is continuous.[19]

*4.1.3 The Parallel Trend Assumption*

To ensure the DID specification captures the treatment effect properly, the outcome variable $OFDI_{ijt}$ must exhibit a parallel trend between treated and control firms prior to the policy intervention (i.e., the parallel trend assumption). To test this, following Shi and Zhang (2023), we estimate the following fixed effects model: $OFDI_{ijt} = \tilde{\beta}_0 + \sum_{\tau=2}^{T} \tilde{\beta}_{1,\tau} \mathbb{1}(t = \tau) \times dT_i + x'_{ijt}\tilde{\beta}_2 + \nu_{ij} + \varepsilon_{ijt}$, where $\mathbb{1}(\cdot)$ is the indicator function that takes a value of 1 if its argument is true. If the parallel trend assumption holds, the estimated $\tilde{\beta}_{1,\tau}$ should not be either individually or jointly statistically significant prior to the policy implementation.

*4.1.4 Robustness Check*

In estimating Equation (5), we will adopt a build-up approach. Initially, we will start with a baseline specification that includes no control variables. We will then incrementally add control variables until the full specification is reached. A robust treatment effect estimate should exhibit minimal variation in the point estimate of $\beta_2$ across these different specifications.

---

[18] However, it remains unclear why this should be the case in reality.

[19] Note that a potential drawback of this approach is that, if $x_{ijt}$ is endogenous—correlated with unobserved factors in the error term—its estimated coefficients may contradict theoretical expectations, as they pick up the confounding effects that operate through correlation with the unobserved factors. However, our objective is not to accurately estimate the marginal effect of $x_{ijt}$.

The policy was motivated by the pre-treatment (before 2017) pollution trend, which can be correlated with paper manufacturing firms' decision on vertical OFDI. For example, observing the pollution trend, firms may foresee the government's regulatory measures, and decide to conduct vertical OFDI. To address this potential confounding effect, we follow the Li, Lu & Wang (2020) to explicitly control for the effect of these pre-treatment pollution trend in the robustness check. Specifically, we compute the average level and changes of $SO_2$ emission two years before the policy (i.e. 2015 and 2016), interact them with a fourth-order polynomial in time, and include them into the regression.

In addition, if the government implemented other policies around the same time as the waste paper import ban that could influence firms' decisions regarding vertical OFDI, their effects might be inadvertently captured in our estimates. However, this concern is mitigated for two reasons. First, the unique characteristics of the paper manufacturing supply chain allow us to plausibly isolate the effect of the waste paper import ban. Second, we conducted a thorough review of news sources and official government policy websites and found no evidence of contemporaneous policies that would confound our analysis.

*4.2 Aggregate-Level Analysis*

As an alternative to the firm-level DID regression model specified in equation (5), we can first estimate the probability of vertical OFDI non-parametrically for both the treated and control groups using firm-level data. The probability of vertical OFDI for group $j$ is estimated as:

$$\widehat{Pr}(OFDI_{jt} = 1) = \frac{1}{|\mathcal{F}_{jt}|}\sum_{i \in \mathcal{F}_{jt}} OFDI_{ijt} \tag{6}$$

where $\widehat{Pr}$ represents the probability estimate, and $\mathcal{F}_{jt}$ denotes the set of firms in group $j$ (treated or control) in year $t$. This is a relative frequency estimator of the probability of vertical OFDI, which converges to the true probability as the number of firms $|\mathcal{F}_{jt}|$ increases to infinity, by the Weak Law of Large Numbers (WLLN).

Once we have the estimated probabilities of vertical OFDI, we can compare the treated and control groups before and after the policy change, following a DID approach to estimate the treatment effect. For this purpose, aggregate equation (5) over $(j, t)$, we obtain:

$$\widehat{Pr}(OFDI_{jt} = 1) = \tilde{\beta}_0 + \beta_1 dP_t + \beta_2 dP_t \times dT_j + x'_{jt}\beta_3 + v_j + \varepsilon_{jt} \tag{7}$$

where $x_{jt} = \frac{1}{|\mathcal{F}_{jt}|}\sum_{i \in \mathcal{F}_{jt}} x_{ijt}$ represents the within-group average of the control variables; $v_j$ is the group-specific fixed effect; $\tilde{\beta}_0 = \beta_0 + E[v_1]$, where $E[v_1]$ is the mean of $v_i$; and $\varepsilon_{jt} = \frac{1}{|\mathcal{F}_{jt}|}\sum_{i \in \mathcal{F}_{jt}}[\varepsilon_{ijt} + (v_i - E[v_1])]$. Note due to the WLLN, $\varepsilon_{jt}$ shall be small if $|\mathcal{F}_{jt}|$ is large. The other right-hand-side variables, such as $dP_t$ and $dT_j$, are defined in the same way as in equation (5).

While this within-group averaging simplifies the model by collapsing many of the firm-level zeros (firms that do not engage in vertical OFDI), it comes at the cost of reducing the sample size. Nevertheless, it provides a robustness check to the firm-level analysis. Equation (6) is a non-parametric estimate of the probability of vertical OFDI, while equation (7) is a parametric DID design. Combining both forms the basis of the aggregate-level analysis, which in this sense is semi-parametric.

## 5. Data and Variables

Data are firm level data sourced from S&P Capital IQ Pro,[20] and consist of 40 publicly listed firms and 2 private firms over 24 years, which have their headquarters based in mainland China. Firms belong to the paper product industry and paper and plastic packaging products and materials industry.[21] Within these 42 firms, 20 firms are treated (i.e., producing corrugated paper and container board in their product mix). The sample period is 2000-2023, where not all firms appear in the whole sample period. That is, data are an unbalanced panel dataset.

A caveat is that the sample primarily consists of listed firms, which are presumably the top firms in the industry due to the requirements of listing in the stock exchange. Nevertheless, this appears of little harm for our purpose. First, theoretically, this corresponds to the cut-off capability endowment ($\underline{\lambda}$) being higher than the minimal level. However, this does not change the theoretical prediction of the model. Second, empirically, our DID research design is comparing listed firms with listed firms. Hence, we are not estimating the treatment effect by comparing apple with banana.

S&P Capital IQ Pro tracks these firms' overseas merger and acquisition (M&A) history, from which we can construct their number of vertical OFDI projects in each year.[22] An overseas M&A is vertical OFDI if (1) the transaction announcement explicitly states the reason of M&A being upstream and downstream integration or (2) the target firm produces pulp or deals with waste paper. In the appendix, we report the classification of overseas M&A as vertical OFDI in details.

Then the outcome variable, $OFDI_{ijt}$, is constructed accordingly. Figure 7 presents the total number of vertical OFDI projects by treatment status.[23] Three observations emerge. First, investing overseas for cheaper intermediate inputs is rare for the paper product industry. Before 2017, no firm had ever conducted vertical OFDI. Second, after 2017, firms in the treated group clearly increased their overseas vertical investment activities. In contrast, firms in the control group continued to have no vertical investment overseas. Third, despite the post-policy spike, the overseas investment activities tapered off eventually. This occurs due to the nature of cost-saving investment. That is, once a firm invests overseas and achieves savings in the cost of intermediate inputs, it does not need to further invest unless the market condition has changed.

---

[20] https://www.spglobal.com/capital-iq-pro.

[21] Two firms primarily classified within other industries also engage in business within the paper product industry.

[22] Consequently, greenfield investments are excluded from this analysis. Given the relatively small number of overseas M&As reported in Figure 7, it is reasonable to assume that greenfield investments—if any—would be similarly limited in number. To assess this, we conducted an extensive search of publicly available news sources. While we were able to identify several post-policy M&A announcements (see footnote 9), we found no media reports indicating the occurrence on greenfield investments during the same period.

[23] See Appendix 2 for a graph of total OFDI (overseas M&A).

Figure 7. The Number of Vertical Outward FDI (Overseas M&A) Projects

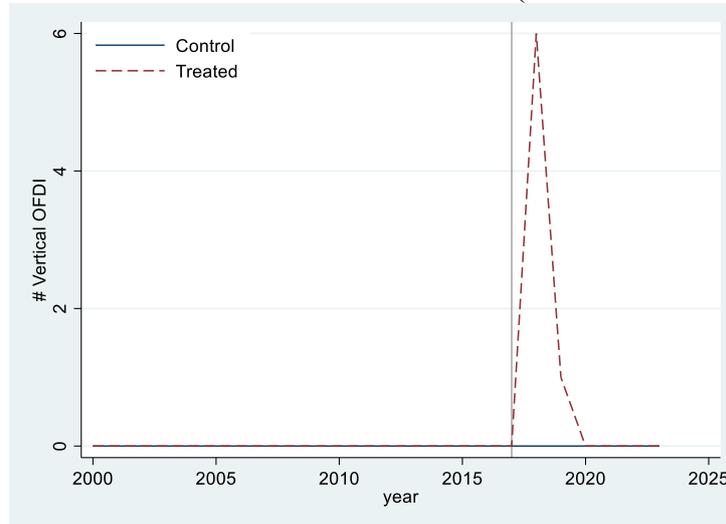

The vector $x$ in equations (5) and (7) consists of firm size, profitability, and age. All else equal, larger firms are more likely to invest overseas because they possess greater financial and organizational resources to overcome the challenges associated with foreign investment. Likewise, more profitable firms are expected to be more inclined towards international expansion, as profitability enhances a firm's internal capacity to invest abroad. Firm age, which proxies for accumulated experience, may also conceptually affect a firm's propensity to engage in overseas investments. However, these firm characteristics may be endogenous in the regressions. To address potential endogeneity, we use a control function approach, where the estimated coefficients can pick up the confounding effects that operate through their correlation with the error term, resulting in deviation from theoretical expectations.

Firm size is measured as the natural logarithm of the ratio of a firm's total revenue to the industry average. Profitability is measured by return on assets (ROA), and age is defined as the number of years since the firm was founded. Total revenue and ROA are directly sourced from the S&P Capital IQ database, and the founding years of the firms are derived from their company profiles.

Table 1. Summary Statistics

| Variable | Obs | Mean | Std. dev. | Min | Max |
|---|---|---|---|---|---|
| OFDI | 725 | 0.0248 | 0.1557 | 0 | 1 |
| firm size | 725 | -0.7191 | 1.2161 | -5.9711 | 2.1177 |
| age | 706 | 30.3683 | 21.6620 | 1 | 100 |
| ROA | 689 | 0.0394 | 0.0358 | -0.0943 | 0.2464 |

Source: S&P Capital IQ Pro.

Table 1 reports the summary statistics of the variables used in the DID regressions. In line with Figure 7, the sample mean of OFDI (the proportion of firm-year observations where the value is 1) is low, at just 2.48%. The average firm size (in natural logarithms) is negative, suggesting the presence of some large firms in the industry, as measured by total revenue. Given the capital-intensive nature of the paper product industry, it is not surprising that larger firms are prevalent. The average firm age is around 30 years, with a standard deviation of 21.662. Profitability within the industry appears modest, with an average ROA of 3.94% and a standard deviation of 0.0358. Overall, the data exhibit sufficient variation for estimating the DID model.

Figure 8. The Distributions of Firm Size, Age and ROA by Treatment Status

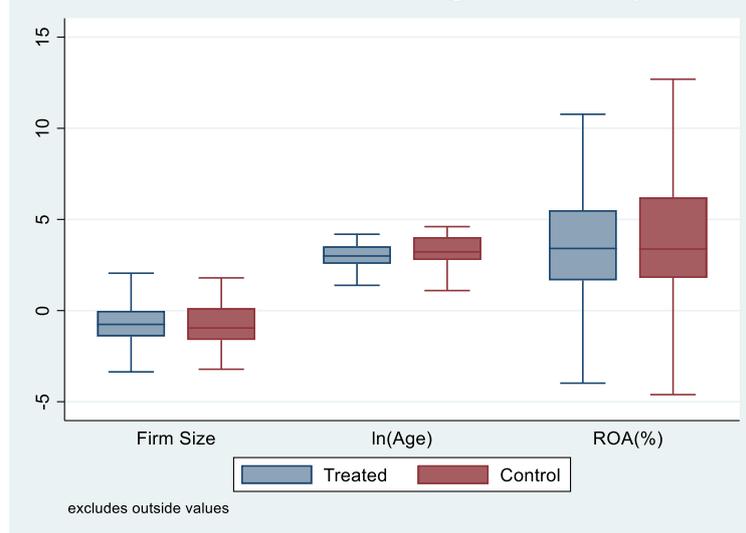

Figure 8 presents the distributions of firm size, age, and ROA by treatment status. While observable differences exist between the treated and control groups, these differences appear to be modest. First, the median values for firm size, age, and ROA in the treated group are approximately equal to those in the control group. Second, although some variation is evident, the interquartile ranges (as reflected in the box sizes) of the treated firms closely resemble those of the control firms. These observations suggest an approximate balance between the two groups in terms of these key firm characteristics.

Importantly, there is no compelling a priori reason to believe that firms in the treated group—those producing corrugated paper and container board—should be systematically different from control firms in the same industry with respect to size, age, or profitability. This visual evidence supports the plausibility of the identification strategy.

**6. Results and Discussion**

In this section, we present the results from the DID regressions at both the firm and aggregate levels. We begin by discussing the baseline model, which excludes control variables, and then progressively introduce additional covariates. The final specification incorporates a second-degree polynomial function of the covariate vector $x$. Throughout all regressions, we use standard errors that are robust to both heteroskedasticity and autocorrelation.

*6.1 Firm-Level Estimates*
The coefficient of the interaction term $dT \times dP$ is of primary interest as it captures the treatment effect of the waste paper import ban. In the baseline specification, the point estimate for this interaction term is 0.1639 with a robust standard error of 0.0365, which is statistically significant at the 1% level. Across all subsequent specifications, the interaction term continues to be statistically significant at the 1% level, with the magnitude of point estimates exhibiting little variation. These results suggest that the waste paper import ban has a significant impact on firms in the paper product industry, prompting them to engage in vertical OFDI.

This finding is consistent with broader industry dynamics. Firms in the paper products sector have traditionally been hesitant to invest abroad. However, the import ban on raw materials such as waste paper has raised — or is perceived to have raised — the cost of intermediate inputs for paper manufacturers. In response, firms are incentivized to engage in upstream vertical OFDI as a strategy to offset these cost pressures. According to estimates from the baseline specification, the policy shock increased the probability of vertical OFDI by 16.39% in the post-policy period relative to the control group.

Table 2. Firm-level DID Regression Results

|  | [1] | [2] | [3] | [4] | [5] |
|---|---|---|---|---|---|
| dP |  | -0.0018 | 0.0017 | -0.0407*** | 0.0049 |
|  |  | (0.0021) | (0.0034) | (0.0156) | (0.0137) |
| dT×dP | 0.1639*** | 0.1567*** | 0.1650*** | 0.1753*** | 0.1650*** |
|  | (0.0365) | (0.0346) | (0.0360) | (0.0378) | (0.0289) |
| firm size |  | 0.0342** | 0.0366** | 0.0354** | -0.0456* |
|  |  | (0.0137) | (0.0150) | (0.0142) | (0.0237) |
| ROA |  |  | 0.1890 | 0.3349* | -0.4668 |
|  |  |  | (0.1712) | (0.2004) | (0.6748) |
| age |  |  |  | 0.0045*** | -0.0072*** |
|  |  |  |  | (0.0016) | (0.0022) |
| (firm size)$^2$ |  |  |  |  | 0.0414*** |
|  |  |  |  |  | (0.0063) |
| ROA$^2$ |  |  |  |  | 1.6705 |
|  |  |  |  |  | (1.5978) |
| age$^2$ |  |  |  |  | 0.0001*** |
|  |  |  |  |  | (0.0000) |
| (firm size)×ROA |  |  |  |  | 0.0279 |
|  |  |  |  |  | (0.3067) |
| (firm size)×age |  |  |  |  | 0.0050*** |
|  |  |  |  |  | (0.0008) |
| ROA×age |  |  |  |  | -0.0007 |
|  |  |  |  |  | (0.0079) |
| Observations | 725 | 725 | 689 | 672 | 672 |
| R$^2$ | 0.0864 | 0.1690 | 0.1803 | 0.1973 | 0.4144 |
| # firms | 42 | 42 | 42 | 41 | 41 |

Note: [1] The inclusion of *dP* results in a non–full-rank estimated covariance matrix of the moment conditions and is therefore partialled out; Heteroskedasticity and autocorrelation robust standard errors in parentheses; *** $p < 0.01$, ** $p < 0.05$, * $p < 0.10$

Turning to the control variables, it is important to note that these may be subject to endogeneity, potentially capturing the confounding effects that operate through correlation with the error term. Nevertheless, the point estimates for these variables, as reported in Table 2, appear to fall within a reasonable range.

Firm size, for example, has a positive and statistically significant marginal effect in columns [2]-[4]. Although the coefficient estimate for firm size is negative and statistically significant at the 10% level in column [5], the squared term and its interaction with firm age have positive and statistically significant coefficient estimates.

Regarding profitability (measured by ROA), the coefficient estimate is not statistically significant in columns [3] and [5], but it is positive and significant at the 10% level in column [4]. Lastly, firm age has a positive and statistically significant coefficient estimate in column [4], while it is significantly negative and its squared term and interaction with firm size have significantly positive coefficients in column [5]. Therefore, experience plays an important role in firms' vertical outward investment.

Overall, these results provide a robust support for the hypothesis that the waste paper import ban has driven firms to pursue vertical OFDI, and the control variables offer additional insights into the characteristics that influence a firm's likelihood of engaging in such investments.

The DID regression fundamentally relies on the parallel trend assumption, which posits that in the absence of the treatment, the treated and control groups would have followed similar trends over time. Figure 9 illustrates the test for this assumption, with the maroon curve connecting each point from the years 2001 to 2023, where the points represent the point estimates of the interaction terms between year dummy variables (with 2000 as the base year) and the treatment status dummy variable. The accompanying error bars indicate the 95% confidence intervals for these estimates.

According to the parallel trend assumption, we expect to see that the coefficient estimates for the period prior to the implementation of the policy are not statistically significant. Figure 9 confirms this expectation. In addition, a $\chi^2$ test for joint significance of the pre-policy interaction terms obtain a test statistic of 6.08 (*p*-value: 0.9783). This result lends credence to the validity of the parallel trend assumption.

Figure 9. The Parallel Trend Assumption

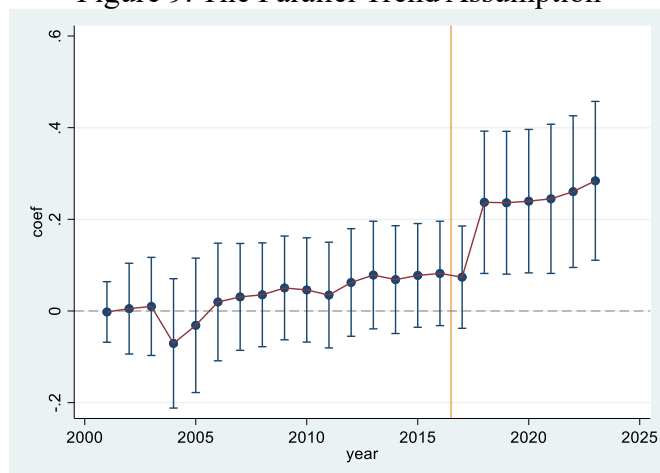

Note: the pre-policy coefficient estimates are also not jointly statistically significant.

Regarding the potential confounding effect of the pre-treatment pollution trend, as described in subsection 4.1.4, we explicitly control them. In the regression, we find a coefficient estimate of 0.1599 (robust standard error 0.0281) for the interaction term *dT×dP*, in line with those reported in Table 2. Therefore, the pre-treatment pollution trend is not a concern for our treatment estimate.

*6.2 Aggregate Level Estimates*

Table 3 presents the aggregate level DID-FE estimation results. The coefficient estimates for the interaction term (*dT×dP*) are positive and statistically significant at the 1% level across all specifications. The consistency of these statistically significant estimates reinforces the conclusion that the waste paper import ban in China indeed encourages firms in the paper product industry to pursue overseas investments in the upstream sub-industry, thereby seeking to mitigate the cost increase of intermediate inputs.

When comparing the estimates from the aggregate level analysis to those obtained from the firm level regressions, we observe minimal variation in the treatment effect estimates, further confirming the robustness of our findings. In contrast, the coefficient estimates for the control variables exhibit some variation between Tables 2 and 3, suggesting that the within-group averaging of $x$ influences the estimation results. Notably, the coefficient estimates for control variables in Table 3 become largely not statistically significant. Intuitively, the within-group averaging of $x$ smooths out the marginal effects of $x$, given the sufficiently large number of firms (22 for the control group and 20 for the treated group). Nevertheless, the results from both Tables 2 and 3 consistently indicate a significant positive treatment effect resulting from waste paper import ban policy.

Table 3. Aggregate Level DID Regression Results

|  | [1] | [2] | [3] | [4] | [5] |
|---|---|---|---|---|---|
| dP |  | -0.0009 | -0.0038 | -0.0041 | 0.0062 |
|  |  | (0.0008) | (0.0051) | (0.0054) | (0.0051) |
| dT×dP | 0.1362*** | 0.1370*** | 0.1375*** | 0.1371*** | 0.1264*** |
|  | (0.0222) | (0.0220) | (0.0166) | (0.0160) | (0.0167) |
| firm size |  | -0.0028 | 0.0065 | 0.0064 | -0.1164 |
|  |  | (0.0025) | (0.0096) | (0.0097) | (0.0731) |
| ROA |  |  | -0.9460* | -0.9346 | 0.1811 |
|  |  |  | (0.5650) | (0.6020) | (2.4424) |
| age |  |  |  | 0.0001 | 0.0052 |
|  |  |  |  | (0.0010) | (0.0065) |
| (firm size)$^2$ |  |  |  |  | 0.0036 |
|  |  |  |  |  | (0.0351) |
| ROA$^2$ |  |  |  |  | -14.0040 |
|  |  |  |  |  | (26.1438) |
| age$^2$ |  |  |  |  | -0.0000 |
|  |  |  |  |  | (0.0001) |
| (firm size)×ROA |  |  |  |  | 0.1252 |
|  |  |  |  |  | (1.2689) |
| (firm size)×age |  |  |  |  | 0.0043 |
|  |  |  |  |  | (0.0030) |
| ROA×age |  |  |  |  | -0.0038 |
|  |  |  |  |  | (0.0701) |

| Observations | 48 | 48 | 48 | 48 | 48 |
| --- | --- | --- | --- | --- | --- |
| $R^2$ | 0.6695 | 0.8022 | 0.8486 | 0.8486 | 0.8623 |

Note: [1] The inclusion of *dP* results in a non–full-rank estimated covariance matrix of the moment conditions and is therefore partialled out; Heteroskedasticity and autocorrelation robust standard errors in parentheses; *** $p < 0.01$, ** $p < 0.05$, * $p < 0.10$

## 7. Conclusion and Implications

This paper investigates the impact of a quasi-experimental policy shock—the 2017 waste paper import ban in China—on firms' likelihood of engaging in vertical outward foreign direct investment (OFDI), combining theoretical modelling with empirical analysis. Within a standard framework of heterogeneous firms competing monopolistically in the final goods market, we show that when the relative unit cost of intermediate inputs (domestic versus foreign) is below one, no firms engage in vertical OFDI. However, when this cost ratio exceeds one, a sorting pattern emerges whereby more productive firms choose vertical integration abroad, while less productive firms do not. The imposition of a raw material import ban disrupts input supply chains, increasing the equilibrium cost of intermediate inputs. If this increase is sufficiently large, it triggers firms' incentive to pursue vertical outward investment.

To test these theoretical predictions, we employ a difference-in-differences (DID) estimation strategy using firm-level data from China's paper product industry. Two features support the identification strategy. First, the waste paper import ban constitutes a quasi-natural experiment—an environmental regulation not explicitly targeting firms' foreign investment behavior. Second, the specific structure of the paper manufacturing supply chain—where waste paper pulp is used exclusively for producing corrugated paper and container board—allows us to clearly distinguish treated firms from untreated counterparts.

Our DID estimates reveal a positive and statistically significant treatment effect: firms in the treated group experienced an approximately 16% higher probability of engaging in vertical OFDI after the policy was implemented, compared to control firms. This result offers firm-level evidence that complements existing macro-level findings, demonstrating that environmental regulations can catalyze international expansion strategies, especially through vertical FDI channels. Moreover, the findings elucidate the underlying mechanism by which regulatory shocks reshape firm-level global strategies.

The findings of this paper carry several policy and managerial implications. First, the observed increase in vertical OFDI highlights an important unintended consequence of environmental regulations: they may incentivize firms to reconfigure global value chains through overseas investment. Policymakers should anticipate such strategic responses when designing regulatory frameworks. Second, firms operating in similarly constrained input markets may view such regulatory disruptions not merely as risks, but as potential catalysts for internationalization. This underscores the role of adaptability, resource endowment, and strategic foresight in managing regulatory uncertainty.

Our findings also point to fruitful directions for future research. The heterogeneous effects of environmental policies across industries and institutional contexts remain underexplored. Expanding the empirical scope to other sectors and countries would help to generalize our insights and uncover additional pathways through which regulation shapes global investment behavior. Finally, our study contributes to the literature on the micro-foundations of FDI by incorporating regulatory shocks into firm-level decision-making within a monopolistically competitive setting. This integration opens the door to richer theoretical models and empirical analyses that link regulatory change, firm heterogeneity, and international strategy.

**Appendix**

**Appendix 1. The Classification of Overseas M&A as Vertical OFDI**

| ID | Year | Ultimate Buyer | Target Firm | Target Firm Product | Country/Region | Vertical OFDI |
|---|---|---|---|---|---|---|
| 1 | 2008 | Nine Dragons Paper (Holdings) Limited | Cheng Yang Paper Mill Co., Ltd. | Industrial papers, including kraft paper, testliner, and medium paper | Vietnam | No |
| 2 | 2017 | Hengan International Group Company Limited | Wang-Zheng Berhad | Disposable fibre-based products, processed papers, wood-free and art papers and consumer and household items | Malaysia | No |
| 3 | 2017 | Shanying International Holdings Co., Ltd | Nordic Paper Holding AB (publ) | Natural greaseproof and kraft paper | Sweden | No |
| 4 | 2018 | Nine Dragons Paper (Holdings) Limited | Bleached Kraft Pulp Mill of CVG, Inc. | Pulp | USA | Yes[1] |
| 5 | 2018 | Nine Dragons Paper (Holdings) Limited | Recycled Pulp Mill of Fibrek Recycling U.S. Inc. | Pulp | USA | Yes[2] |
| 6 | 2018 | Nine Dragons Paper (Holdings) Limited | Catalyst Paper Operations Inc. | Pulp and paper | USA | Yes |
| 7 | 2018 | Shandong Sunpaper Co., Ltd. | Birla Lao Pulp & Plantation Co., Ltd. | Pulp and eucalyptus tree plantation | Laos | Yes |
| 8 | 2018 | Shanying International Holdings Co., Ltd | Verso Wickliffe LLC | Pulp and kraft packaging paper | USA | Yes[3] |
| 9 | 2018 | Shanying International Holdings Co., Ltd | Waste Paper Trade C.V. | Various types of recovered materials, including paper, plastics, and film | Netherlands | Yes |
| 10 | 2018 | Xiamen Hexing Packaging Printing Co., Ltd. | United Creation Packaging Solutions (Asia) Pte. Ltd. | An investment holding company, corrugated Box | Singapore | No[4] |
| 11 | 2019 | Nine Dragons Paper (Holdings) Limited | Wiseland International Holdings Limited | Unknown | Hong Kong | No[5] |
| 12 | 2020 | Shandong Chenming Paper Holdings Limited | VNN Holdings Limited/Aberdeen Industrial Limited | An investment holding company | Hong Kong | No |
| 13 | 2021 | Nine Dragons Paper (Holdings) Limited | Turbo Best Holdings Limited | Corrugated paperboard and paper-based packaging products | Hong Kong | Yes[6] |

Note: An overseas M&A is vertical outward FDI if (1) the transaction announcement explicitly states the reason being upstream and downstream integration or (2) the target firm produces pulp or deals with waste paper;
[1] https://us.ndpaper.com/nd-paper-completes-acquisition-of-old-town-maine-pulp-mill/;
[2] https://us.ndpaper.com/nd-paper-completes-acquisition-of-recycled-pulp-mill-in-west-virginia/;
[3] https://www.recyclingtoday.com/news/paper-mill-kentucky-global-win-former-verso-kraft/;
[4] Horizontal integration, https://www.capitaliq.spglobal.com/apisv3/docviewer-service/go/9rAVLTVd;
[5] Business combination involving entities under common control, https://www.capitaliq.spglobal.com/apisv3/docviewer-service/go/meuwIlVd;
[6] Upstream and downstream integration, https://doc.irasia.com/listco/hk/ndpaper/announcement/a249604-eng_ann_acquisition2.7.2021.pdf.
Source: Capital IQ Pro.

## Appendix 2. The Number of Total OFDI (Overseas M&A) Projects

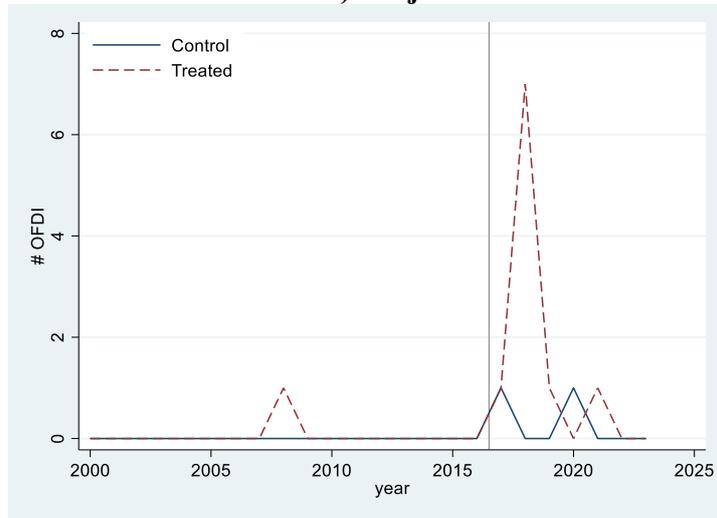